\documentstyle[epsf,fleqn,12pt]{article}
\begin{document}
\baselineskip=18pt
\parskip=3pt

$\phantom{}$ \vskip 1cm

\begin{center}
{\LARGE Growth diversity in one dimensional 
fluctuating interfaces}
\end{center}

\vskip 0.5cm

\begin{center}
{\large M. D. Grynberg}
\end{center}
\begin{center}
\large{Departamento de F\'{\i}sica, 
Universidad Nacional de La Plata\\
C.C.  67, (1900) La Plata, Argentina}
\end{center}

\vskip 1cm

\begin{center}
{\bf Abstract}
\end{center}

A set of one dimensional interfaces involving attachment and
detachment of $k$-particle neighbors is studied numerically using 
both large scale simulations and finite size scaling analysis. 
A labeling algorithm introduced by Barma and Dhar in related spin 
Hamiltonians enables to characterize the asymptotic behavior of the 
interface width according to the initial state of the substrate. 
For equal deposition--evaporation probability rates it is found that 
in most cases the initial conditions induce regimes of saturated width. 
In turn, scaling exponents obtained for initially flat interfaces indicate 
power law growths which depend on $k$. 
In contrast, for unequal probability rates the interface width exhibits a 
logarithmic growth for all $k > 1$ regardless of the initial state
of the substrate.

\vskip 1cm



Published in  {\sc J. Stat. Phys. {\bf 103}, 395 (2001)}.
\newpage

Studies of interface growth have acquired 
a major momentum in the past decade \cite{Krug,HZ,Meakin}, 
finding a rich variety of applications. 
These range from molecular beam epitaxy \cite{Politi} 
to magnetic flux lines in type-II superconductors \cite{Krug,Hwa}
and polymer physics \cite{Kardar}.
Most numerical analysis and theoretical investigations
on  growing structures have emphasized the
onset of scaling regimes which emerge at both large time
and length scales. This enables a classification
of nonequilibrium  interfaces
by universality classes characterized by a set 
of scaling exponents which, independently of the
microscopic growth rules, dominate the late evolution 
stages of these processes. 

However, the issue of universality classes in these problems
has been steadily brought into question, particularly in higher 
dimensions \cite{Krug2,Newman,Lopez,Koduvely}.
Since no systematic analytical method is yet available
to understand large scale fluctuations in nonequilibrium systems, 
two main research lines have been addressed to probe the
universality hypothesis. They involve either the simulation
of discrete models \cite{Koduvely,Kertesz,Ala-Nissila,Forrest,Meakin2} 
or the numerical solution of phenomenological Langevin-type 
equations of growth driven by Gaussian noise \cite{Amar}. 
This is the case of the ubiquitous Kardar-Parisi-Zhang (KPZ) equation 
\cite{KPZ} which has been applied to characterize a vast body of
far from equilibrium processes \cite{Krug,HZ,Meakin}.

In this letter we rather follow the first category of investigations
and  introduce a set of restricted solid-on-solid (RSOS)
growth models in one dimension entailing an exceptionally large number of
conservation laws \cite{Barma}. By mapping the problem onto a 
quantum spin system, it will turn out that for equilibrium situations,
not only each model defines a ``class'' of its own but furthermore, 
their asymptotic properties become extremely sensitive to slight deviations 
from initially flat substrates. In contrast, non-equilibrium situations 
are rather robust but exhibit a much slower dynamics.
This will mark a clear difference between fluctuations 
in equilibrium and non-equilibrium cases. 

The basic process considered is a simple yet nontrivial extension
of ballistic aggregation models \cite{Meakin2}. Here, we include
both deposition and evaporation of extended objects such as
dimers, trimers, ... $k$-mers, at random locations of a 
one-dimensional interface \cite{Hinrichsen}.
As usual, the state of the system at a given time 
is defined by a set of single valued functions 
$\{h_1 (t),\,...\,,h_L (t)\}$, where $h_n$ is the
height of the interface measured from a reference level
at site $n \in \{1,2, ...\,,L\,\}$. To prevent the divergence
of surface fluctuations at large times we also impose
a RSOS constraint  throughout the process namely,  
$\vert\, h_{n+1} - h_n\, \vert \equiv 1\,,\,\, \forall\, t\,$.
Therefore, a $k$-mer deposition (evaporation) attempt  with
rate $\epsilon\,$ ($\epsilon'\,$) is successful provided 
a plateau with at least  $k$-successive
local minima (maxima) is selected from the interface. 
Fig. 1 illustrates these microscopic rules for the case of dimers.
Notice that evaporation apply whether or not the targeted
$k$-adjacent maxima were created together, a feature which   
allows for reconstitution of $k$-mers. 
Also, it is helpful pointing out that the dynamics of these 
interfaces is equivalent to that defined in the $2k$-mer 
deposition-evaporation problem studied in Ref. \cite{Barma},
by flipping the spins on all even sites, say, of this latter system.

In studying the scaling regimes of growing structures
it is useful to consider the mean square fluctuations $W^2$
of the average interface height $\langle h (t) \rangle\,$, i.e.
$W^2 (L,t) = \frac{1}{L} \, \sum_j \langle\, \left[\, h_j(t) -
\langle\, h(t)\,\rangle\,\right]^2\,\rangle\,$, which yields
a measure of the interface width. 
From the assumption of a scaling 
form of height-height correlation functions along
with a single time-dependent length scale 
$\sim t^{1/z}$ for growth starting from an initially flat interface, 
it can be argued \cite{Family} that $W$ scales as
\begin{equation}
\label{width}
W (L,t) = L^{\zeta} \,f(t/L^z)\,,
\end{equation}
where the scaling function $f (c)$ satisfies
\begin{equation}
\label{scaling}
f (c) \sim \cases{ c^{\,\zeta/z} \:\:\:\:\:\:\:
{\rm for} \:\:\: c \ll 1\,,\cr
{\rm const} \:\:\:\:\:\, {\rm for} \:\:\:c \gg 1\,. \cr}
\end{equation}
It  follows immediately that finite systems
saturate as $W \propto L^{\zeta}\,$, whereas in the thermodynamic limit 
the asymptotic growth is dominated by the exponent $\beta = \zeta/z$,
that is $W \propto t^{\beta}\,$.  The exponent $\zeta$ describes
the roughness dependence of the surface width on the typical 
substrate size. In turn the exponent $z$, often known
as the dynamic exponent, gives the fundamental scaling between length and time.

However, already for $k \ge 2\,$ it will turn out that there are initial 
substrates whose asymptotic widths neither follow a power law nor exhibit the 
universal characteristics embodied in  Eqs.\,(\ref{width}) 
and (\ref{scaling}). As we shall see, the investigation of these issues
becomes rather systematic in the language of the spin systems discussed below.

\vskip 0.2cm
{\it a. Spin representation} ---
We now exploit the well known mapping between RSOS interface dynamics
and quantum spin systems \cite{Meakin2}.
Associating the height difference $s_n \equiv h_{n+1} - h_n\,$
to an eigenvalue of the $z$-component, say, 
of the Pauli operator $\vec \sigma_n$ for site $n$, then
all intrinsic properties of the interface can be
analyzed in terms of direct products  $\vert s \rangle = 
\vert s_1,\,...\,,s_L \rangle\,$ of 1/2 spinors.
By construction it is clear that up to  a constant,
the interface heights are obtained as 
$h_n = h_1 + \sum_{j=1}^{n-1}\, s_j\,$,
where $h_1 - h_L = s_1 \,$ upon imposing periodic boundary conditions (PBC).
Note that the latter force a vanishing total magnetization 
throughout the underlying spin kinetics.

Introducing spin-$1/2$ raising and lowering operators
$\sigma^+_n$, $\sigma^-_n\,$, the stochastic evolution 
of our $k$-mer interfaces at subsequent times can be calculated 
from the action $e^{-H\,t}\,$ of the quantum ``Hamiltonian''
\begin{eqnarray}
H &=&\sum_j \!\left(\,
\epsilon\,  A^{\!\dag}_j +
\epsilon'\,  A_j \,\right) 
\left(\, A^{\!\dag}_j +  A_j  - 1\right)\,,\\
A^{\!\dag}_j &=& \prod_{i=1}^{k} 
\sigma^+_{j+2i-2}\,\sigma^-_{j+2i-1}\,.
\end{eqnarray}
Here, adsorption (desorption) of $k$-mer plateaus with probability 
$\epsilon\,$ ($\epsilon'$) is described by the effect of the
$A^{\!\dag}\,$ ($A\,$) operators.
Conservation of probability requires in turn the appearance of
$2 k$-field correlators, already diagonal 
in the $\sigma^z$ or particle (monomer) representation.
We address the reader to Ref. \cite{Gunter} for a more
detailed derivation in related systems. 

\vskip 0.2cm
{\it b. Conservation laws} --- The advantage of the spin representation
is that one can identify a large number of subspaces which are 
mutually disconnected by the $H$-dynamics. Following Ref.\cite{Barma},
we define accordingly a reduction rule by looking for the occurrence
of groups of $2k$-opposite {\it consecutive} spins in a given state 
$\vert s \rangle$.  Each occurrence, if any,
is deleted so the length of the remaining spin configuration
is reduced in $2k$ bits per deletion. This procedure is applied 
recursively, until we are left with a string that cannot be further
reduced. Such an object corresponds to an {\it irreducible string}\,
$I \vert s\rangle$ \cite{Barma}. For example, in the dimer case
we have $I \vert\uparrow\uparrow\downarrow\downarrow
\uparrow\downarrow\downarrow\uparrow\rangle =\, 
\uparrow\uparrow\downarrow\downarrow
\uparrow\downarrow\downarrow\uparrow\,$ (irreducible block of Fig. 1),
$I \vert\uparrow\downarrow\uparrow\downarrow 
\uparrow\downarrow\downarrow\uparrow\rangle =\;
\uparrow\downarrow\downarrow\uparrow\,,$ 
$I\vert\uparrow\downarrow\uparrow\downarrow 
\uparrow\downarrow\uparrow\downarrow\rangle = \emptyset\,,$
the null string. 

The key observation is that so long as 
$\epsilon,\,\epsilon' > 0\,$, two spin configurations
$\vert s \rangle\,$ and $\vert s' \rangle\,$ belong to the same
subspace $\iff I \vert s \rangle = I \vert s' \rangle\,$
\cite{Barma}.
Thus, treated as a dynamical variable, the irreducible string
is a constant of motion and provides a unique label for each
$H$-invariant subspace. To illustrate this point consider a sector
labeled by $\chi_1\,,...\,,\chi_{_{\cal L}}$ with ${\cal L} < L$
irreducible characters $\chi =\pm 1 \,$. Let $\{x_j (t)\}$ denote
their positions in the full $L$-site lattice before applying the
reduction rule. The interface dynamics then induces a random
walk of these characters such that $x_{j+1} (t) > x_j(t)\,\, 
\forall\, t\,$. Since $H$ changes $x_j$ by multiples of 
$2k$ lattice spacings, the spin array between sites 
$x_j$ and $x_{j+1}$ can be ultimately reduced to a null string.
Thus, we see that $\{x_j (t)\}$ constitute the slow modes of 
our fluctuating interfaces whereas their characters
remain unaltered throughout the dynamics.
The above reduction algorithm also enables to count the total 
number of invariant sectors, both jammed (strings which are already 
irreducible), and unjammed. A straightforward analysis given as in 
\cite{Barma} shows that these sectors increase as fast as 
$\lambda^L$ for large $L$, where $\lambda$  is the largest root of 
$\lambda^{2 k} = 2 \,\lambda^{2 k - 1} - 1\,$.

\vskip 0.2cm
{\it c. Saturated regimes} --- It is then natural to ask whether 
the dynamics is affected by this rather unusual partition of the 
full phase space. 
At least in the simpler case, when $\epsilon = \epsilon'$ 
\cite{equilibrium}, it can be argued that $W$ {\it remains bounded}
whenever we start from a substrate of finite width and the
length ${\cal L}$ of the corresponding irreducible string is a 
{\it finite} fraction of the lattice size. Specifically, consider
two arbitrary points $A$ and $B$ in a ring of $L$ sites. The most
general configuration of spins determining the height difference
$h_A - h_B\,$  can be denoted as
$$u_1\left\vert^{^{^{\phantom{.}}}} \right. ^{^{^{^{\!\!\!\!\!\!\!h_A}}}}
\!\!\!x_1\,w_1\,x_2\,w_2\;... \; x_{n-1}\,w_{n-1}\,x_n
\left\vert^{^{^{\phantom{.}}}} \right. ^{^{^{^{\!\!\!\!\!\!\!h_B}}}}
\!\!\!u_2\,,$$ 
where $x_i$ are the positions of the irreducible characters
between $A$ and $B$, $w_i$ are spin patches reducible to null strings, 
and $u_1$ ($u_2$) is the part of the irreducible string between the 
previous (next) irreducible character $x_0$ ($x_{n+1}$\,) that lies 
to the left of $A$ (right of $B$). To calculate the height difference
between $A$ and $B$, we can ignore all $w_i$ patches as they
necessarily have equal number of up and down steps. On the other hand,
if the initial substrate has bounded heights, the sequence 
$x_1(t)\,...\: x_n(t)$ can only give a bounded contribution
to the height difference $\forall \,t$. The remaining contributions
at subsequent times comes from $u_1$ and $u_2$. If $u_1$ is part
of a reducible substring  of length $2m$ (between $x_0$ and $x_1$),
the maximum height difference it can have is at most $m$. Also,  
for $\epsilon = \epsilon'$ the detailed balance condition holds
in the steady state.
For such equilibrium situation it has been shown that $m$
has an exponential distribution, provided
that the density ${\cal L}/L$ of irreducible characters is
held finite \cite{Barma}.
Consequently, $u_1$ yields only a finite variance. Similar
considerations apply to $u_2$. Hence, the height difference
between two arbitrary points has bounded variance
$\forall \,t$ and $L$. Therefore, we conclude that the width
of the substrate remains finite, or equivalently,  in the
steady state $W$ should exhibit a saturated regime.
However, notice that if the irreducible string is very short, 
i.e. ${\cal L}/L \to 0\,$, fluctuations in the $(x_{j+1} - x_j)$
distances become large and the above reasoning for $u_1,\,u_2$ 
does not hold.

The case $k=2$ is typical and 
already exhibits such sector dependent behavior. 
Thus, we consider only this case in detail.

\vskip 0.2cm
{\it d. Numerical results} --- Armed with this labeling methodology
along with the spin-height mapping discussed above, 
we studied the dimer interface width  $W$ in a number of representative
subspaces. Fig. 2 displays the width behavior  obtained
from Monte Carlo simulations using 
five initial scenarios. These were embedded homogeneously on 
the Neel state of $2.1 \times 10^5$ lattice 
sites with PBC, and correspond to 
(1) the null string, i.e. the usual flat substrate; 
(2)\, $[\uparrow\uparrow\downarrow\downarrow]^{L/60}$ that is,
the irreducible string constructed by repeating the unit cell 
$L/60$ times; (3)\,$[\uparrow\uparrow\downarrow\downarrow]^{L/8}$;
(4)\, $[\uparrow\uparrow\downarrow\downarrow\downarrow
\uparrow]^{L/14}$; 
(5) a random configuration formed by deposition-evaporation of 
monomers on the Neel state. Setting  $\epsilon = \epsilon'$,  
the averages were taken over $100$ histories of $10^5$ steps per height.

Although for $\epsilon = \epsilon'$ the interface of sector (1) 
moves on the average with zero velocity [see lower inset of Fig. 3\,(a)\,], 
alike the monomer case ($k=1$) it can grow to arbitrarily large heights.
In fact, this is consistent with our data which support a power 
law growth $W^2 \propto t^{2\beta}$ 
extended over more than two decades with an exponent $\beta \sim$ 0.12(1)\,.
On the other hand, sectors (3) and (4) clearly 
exhibit the saturated regimes conjectured
above. They fit well with an exponential form 
$W^2 \propto \exp\left(-\,t^{-\alpha}\right)\,$,
as is shown in the semi-log inset of Fig. 2
upon setting  $\alpha \approx 1/4\,$ and 1/3 respectively \cite{stretched},
in sharp contrast with the expectations of Eqs.(\ref{width}) 
and (\ref{scaling}). 
Thus, we see that certain deviations from a perfectly flat substrate
turn into drastic changes at large times. As it was referred to above, 
the latter depend on both the density $\rho = {\cal L}/L$ 
and particular sequence of 
irreducible characters considered \cite{PBC}.

The situation of sectors (2) and (5) is less clear. Here $\rho$\,
is still finite (though  much smaller than the irreducible density
of the previous non null sectors), and consequently a saturated regime 
would be expectable. However, also here the data extend nearly
over two decades following power law growths 
with $\beta \approx 0.07\,$ and $0.04\,$ respectively
(dashed slopes of Fig. 2). One may attempt to interpret
these results in terms of a temporal crossover, although to reach 
an asymptotic saturated regime such a crossover should
be exceptionally slow. A numerical test in favor of this latter
possibility is provided by the average interface height 
$\langle\, h(t)\,\rangle$. Notice that   
given a distribution of heights $P (h_1,\,...\,, h_L\,,t)$,
the mean height can be evaluated as
\begin{equation}
\hskip -1cm
\partial_t \langle h \rangle = \frac{4}{L} \sum_{\vert h \rangle,\,i} 
\epsilon \,P (...\,\overbrace{h_i^{(1)}\!\!,h_i,h_i^{(1)}\!\!,
h_i\,,h_i^{(1)}}^{\rm dimer}...\,,t) -
\epsilon' P (... \,\overbrace{h_i^{(1)}\!\!,h_i^{(2)}\!\!,
h_i^{(1)}\!\!,h_i^{(2)}\!\!,h_i^{(1)}}^{\rm dimer}...\,,t)\,,
\end{equation}
where $h_i^{(j)} = h_i + j\,$, and the overbraces
involve the heights which form a dimer, say
$h_{i-1},\,...\,,h_{i+3}\,$ (see Fig. 1).
For $\epsilon = \epsilon'$, it follows that
$\langle\, h(t)\,\rangle\,$ must saturate to a finite 
value  at large times, as all accessible
height configurations are equally likely in the steady state, 
{\it irrespective} of the sector considered \cite{equilibrium}.
In fact, Fig. 3\,(a) displays saturation regimes in most sectors.
However, it also reveals the existence of a rather huge temporal 
crossover in sector (2) (more than $10^5$ steps), along with a medium one 
in sector (5) (about $10^3$ steps), possibly responsible 
for the unexpected width behavior observed in Fig. 2.
It is important to note that for $\epsilon > \epsilon'\,$ the
average interface height behaves quite differently, 
at least within our accessible time
scales. The latter can grow indefinitely with a non 
zero velocity, as suggested by the asymptotic logarithmic increase 
of $\langle h(t) \rangle\,$ observed in Fig. 3\,(b) for all sectors.

Before discussing such logarithmic behavior (see section {\it e} below),
we complete our study of the case $\epsilon = \epsilon'\,$
and analyze the late growing stages
of higher $k$-mer interfaces, now focusing attention
on initially null strings, the common flat substrate.
After averaging over 100 histories our results indicate 
a slight dependence of $\beta$ on the value of $k\,$ 
(however, see below the variation of $z$),
which nonetheless can be clearly distinguished over 
five step decades, as indicated by the slopes of Fig. 4. 
The measured values of $\beta$ are presented in Table I
for $k=2,4$ and 8. 

To enable a complementary check of these exponents 
we turn to a finite size scaling analysis
of the interface width. The reader's attention is directed to Figs. 5(a)-(c) 
where we display in turn the results of the 
simulations carried out for $k = 2,4\,$ and 8.
Here, the averages were taken over $2 \times 10^4\,$ samples 
employing periodic chains of $L = \,$ 400, 640 and 800 sites. Using the 
estimates of the dynamic and roughening exponents $\{z, \zeta \}$ given
in Table I, we obtained a fair data collapse towards the (squared) 
phenomenological scaling functions
conjectured by Eqs.\,(\ref{width}) and (\ref{scaling}) 
\cite{Family}. As expected, the resulting values
of $\zeta/z\,$ are in reasonable agreement with the previous
$\beta$-set obtained in the thermodynamic limit. 
So we finally see that as long as the deposition-evaporation rates
are held equal, the emerging
scaling exponents are in fact nonstandard and give rise to rather unusual
dynamic length scales $\,t^{1/z}$ which decrease markedly with $k$.

\vskip 0.2cm
{\it e. Logarithmic growth. 
$\epsilon \ne \epsilon'$} --- The logarithmic dependence 
of the average interface height $\langle h(t) \rangle$ 
obtained in Fig. 3(b) at large times,  may be understood by 
means of the following heuristic considerations.
For clarity of argument, assume $\epsilon \gg \epsilon'$. The random
deposition leaves some single spaces between segments occupied by dimers.
This single spaces cannot be filled by subsequent deposition  of dimers,
with the result that the height profile after many depositions
resembles a ``mountain landscape'' where no further depositions are possible.
Such profile is clearly evidenced by the evolution snapshot given
in the upper panel of Fig. 6.

If $\epsilon'= 0$, this configuration cannot evolve any further,
and the configuration is stuck. If $\epsilon'$ is non-zero, one can
evaporate away from one side of the mountain, all the way to the base, and
then redeposit the base by dimer shifting it one unit (right  or
left, whatever is allowed). Then, rebuilding on this side, 
one gets a different profile (dotted line in the lower panel of Fig. 6). 
This rebuilding process leads to sometimes extinction of small mountains 
(which merge with bigger ones), and thus the average  mountain size 
$M(t)$ increases with time. 
The time dependence of $M$ can be estimated as follows. For a single
rearrangement event described above, one needs order $M$ evaporation
events. Hence by standard activation arguments, this rate is
$(\epsilon'/\epsilon)^M$. One needs order $M^2$ such events for total
width to change by $M$ (since change occurs either way). Hence, the time
for the average mountain size to change from $M$ to $2 M$ is of order $M^2
(\epsilon/\epsilon')^M\,$, so  we get 
$t(2 M) \sim M^2 \,(\epsilon/\epsilon')^M$.
On the other hand, from simple geometry both the average mountain 
height and the width $W$ scale as $M$ (i.e. the roughness exponent 
$\zeta$ is 1). Therefore, this yields
\begin{equation}
W(t) \sim \log(t)\:,\:\:\: \epsilon \ne \epsilon'\,.
\end{equation}
These conclusions can be extended for all 
substrates (whatever the sector or irreducible string considered), as
well as  $ \forall \,k > 1$. In fact, the semi-log data displayed
in Fig. 7 corroborates the above 
conjecture for $k = 2, 4,\,$ and 8 starting from empty substrates. 
Also, the upper inset confirms this logarithmic growth for $k=2$ in 
sectors (2)--(5).
A similar one-dimensional model showing logarithmically slow coarsening
was studied in \cite{Evans}.

To summarize, we have studied a family of simple growth
models yet exhibiting highly nontrivial behavior. 
For $\epsilon = \epsilon'$
our finite size analysis indicates that each $k$-mer process
might be regarded as a growth class of its own dominated by
different dynamic exponents $z(k)$. 
The large systems explored  within the time scales of Fig. 4 
yielded growth exponents consistent with the $\beta$ values obtained 
from finite substrates (Fig. 5).
The concept of irreducible string \cite{Barma}
played a crucial role in determining the saturated behavior
entailed by the initial conditions (Fig. 2), as opposed to 
the much simpler dynamics of monomer interfaces. However, the
late dynamic stages of small strings (${\cal L} \ll L\,$) 
involve slow temporal crossovers for which further numerical efforts
will be required to clarify such situations.
Generalizations of this labeling algorithm to higher dimensions are 
clearly desirable though its mere existence appears difficult to elucidate. 
Whether or not an exponential number of conservation laws
show up in $ d > 1$ yet remains an open question.

For the non-equilibrium situation ($\epsilon \ne \epsilon'\,$),
Fig. 7. indicates a rather robust dynamics. Here,  
the interface width grows logarithmically $\forall \,k > 1\,$ 
irrespective of the initial state of the substrate. 
In view of these drastic differences between fluctuations in equilibrium
and non-equilibrium cases, the transient dynamics of regimes
with $\epsilon \approx \epsilon'\,$ still needs further 
investigations.

\vskip 0.5cm

It is a pleasure to thank R.B. Stinchcombe, M. Barma and 
T.J. Newman for helpful comments. Also, the remarkable 
observations made by the referee contributed to improve this work. 
The author acknowledges support of CONICET, Argentina.


\newpage
\begin{table}
\begin{center}

\begin{tabular} {c c c c} \hline
$\:\:\:\:k\:\:\:\:$ & $\beta\:\,
(L \to \infty)\:\:\:\:\:\:\:\:\:\:\:\:\:\:$  & 
$z\:\:\:\:\:\:\:\:\:\:\:\:\:\:$ & 
$\zeta\:\:\:\:\:\:\:\:\:\:\:\:\:\:$ \\ \hline
2 &  $0.12(1)\;\;\;\;\;\;\;\;\;\;$  & $2.5(10)\;\;\;\;\;\;\;\;$   & 
$0.3(1)\;\;\;\;\;\;\;\;$ \\ 
4 &  $0.10(1)\;\;\;\;\;\;\;\;\;\;$  & $2.8(10)\;\;\;\;\;\;\;\;$   & 
$0.3(1)\;\;\;\;\;\;\;\;$ \\ 
8 &  $0.09(1)\;\;\;\;\;\;\;\;\;\;$  & $3.3(10)\;\;\;\;\;\;\;\;$   & 
$0.3(1)\;\;\;\;\;\;\;\;$ \\ 
\hline
\end{tabular}
\end{center}
\caption{Growth exponent $\beta$ measured in large $k$-mer substrates
(Fig. 4), along with finite size scaling estimations of dynamic and 
roughening exponents $z$ and $\zeta$ (Fig. 5), obtained for 
equal deposition--evaporation rates.}
\end{table}


\begin{figure}
\vbox{%
\vspace{-16cm}}
\hbox{%
\hspace{-2.5cm}
\epsfxsize=8.4in
\epsffile{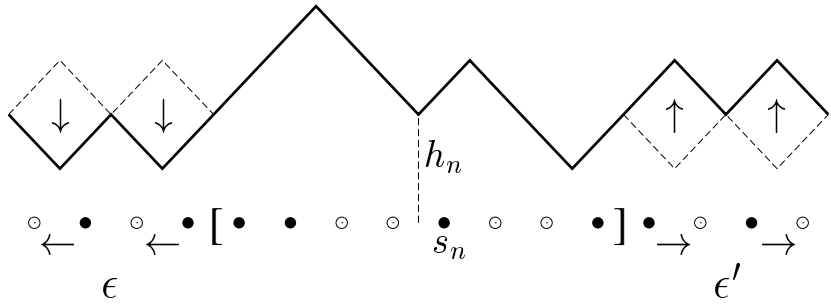}}
\vspace{-8cm}
\caption{Schematic representation of dimer
deposition- evaporation onto a 
RSOS interface. The equivalent  spin-$\frac{1}{2}$ 
($s_n \equiv h_{n+1} - h_n$\,), or hard core particle 
dynamics involves two
left (right) particle hopping at a time
with rate $\epsilon$ ($\epsilon'$) for
dimer adsorption (desorption). The array between
brackets illustrates an irreducible string (see text).}
\end{figure}

\newpage
\begin{figure}
\vbox{%
\vspace{-5cm}}
\hbox{%
\epsfxsize=5.5in
\hspace{0.9cm}
\epsffile{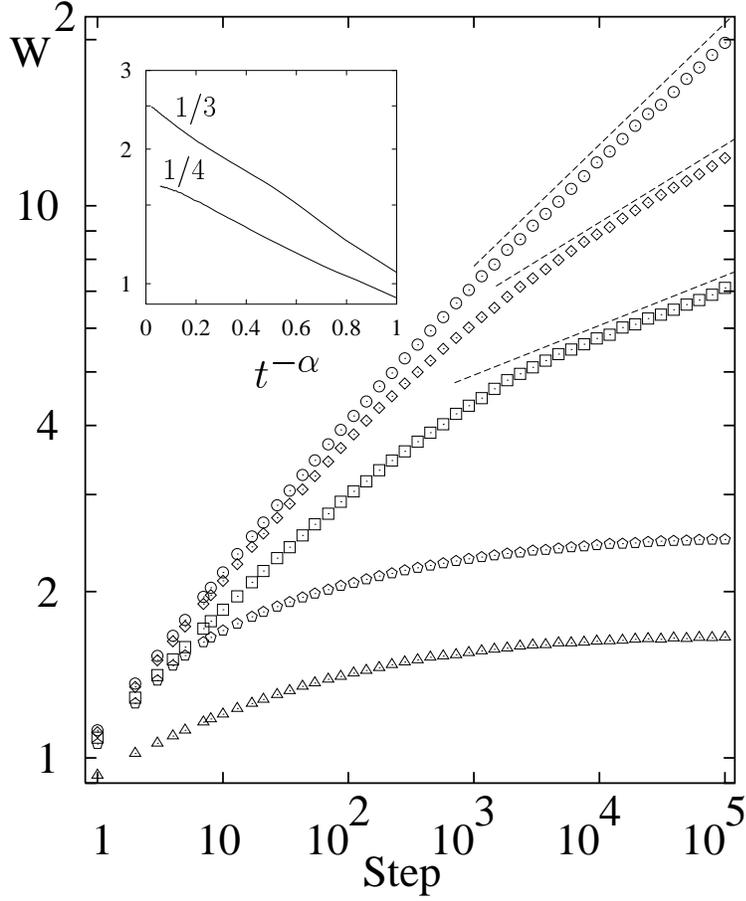}}
\vspace{-1.3cm}
\caption{Evolution of $W^2$ in subspaces (1)-(5) defined
in text. Here, they correspond to (1) circles (null string); (2) romboids; 
(3) triangles; (4) pentagons and; (5) squares.
The slopes of upper straight lines indicate a power-law 
asymptotic behavior $t^{2 \beta}$
with $\beta \approx 0.12$ (top), 0.07 (middle),
0.04 (bottom). These two latter cases however, might be undergoing
a slow temporal crossover (see text).
In turn, subspaces (3) and (4)
fit appropriately an exponential form 
$W^2 \propto \exp\left(-\,t^{-\alpha}\right)$
with $\alpha \approx 1/4$ and 1/3 respectively, 
as shown in the inset.}
\end{figure}

\newpage
\begin{figure}
\vbox{%
\vspace {-10cm}}
\hbox{%
\epsfxsize=5.6in
\hspace{0.05cm}
\epsffile{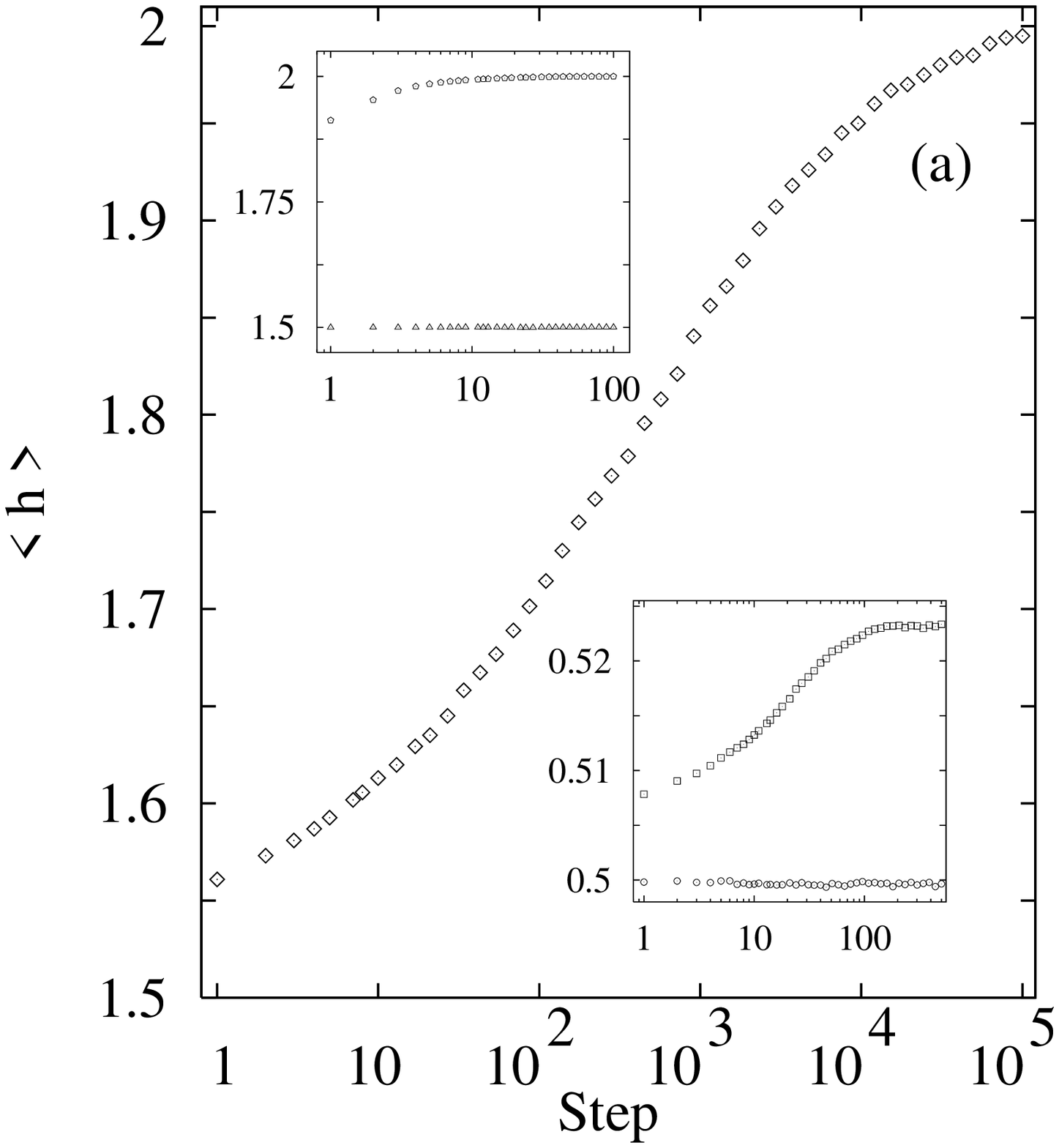}}
\end{figure}

\newpage
\begin{figure}
\vbox{%
\vspace {-1cm}}
\hbox{%
\epsfxsize=4.15in
\hspace{1.1cm}
\epsffile{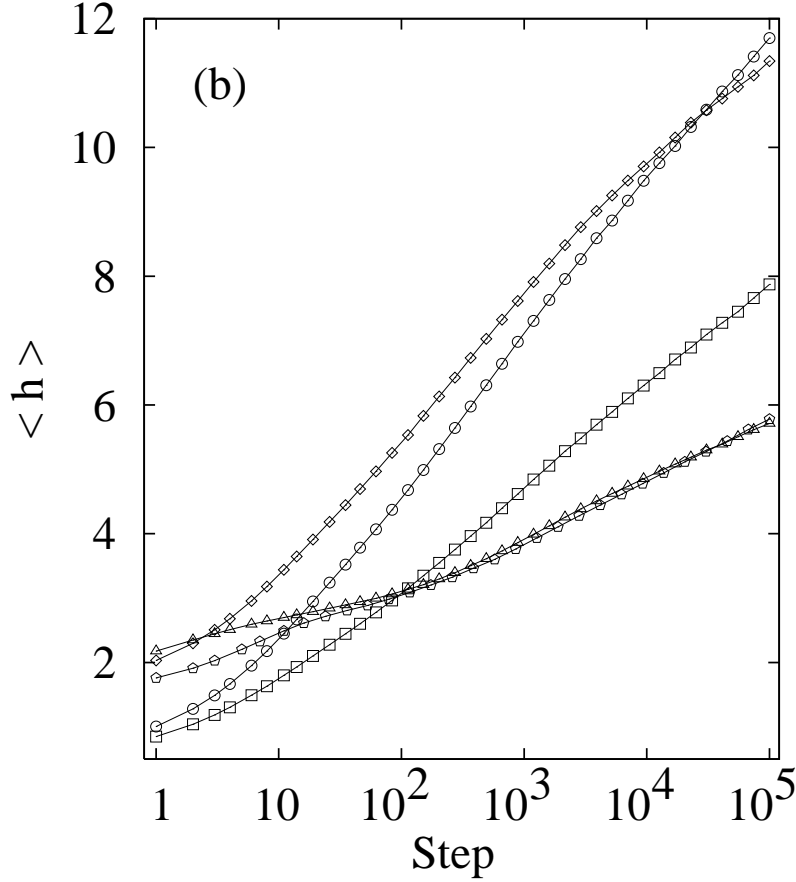}}
\vspace{1.5cm}
\caption{Average interface height $\langle h \rangle$
for $k=2$ with (a) $\epsilon = \epsilon'$, measured from the 
flat interface level  [$ \langle h(0) \rangle= 0.5\,$].
Sectors (1) (bottom curve of lower inset),
(4) and (3) (from top to bottom in the upper inset), display a fast
relaxation toward the expected saturated regime. 
In contrast, sectors (2) (romboids) and (5) (top curve of lower inset), 
undergo a slow temporal crossover. (b) Average height $\langle h \rangle$ 
for $k=2$ with $\epsilon'/\epsilon = 0.4\,$. All sectors exhibit logarithmic
growth at large times as opposed to the saturated regimes in (a).
Solid lines are guide to the eye whereas all symbols are taken as in Fig. 2.}
\end{figure}

\newpage
\begin{figure}
\vbox{%
\vspace {-6cm}}
\hbox{%
\epsfxsize=5.5in
\hspace{0.9cm}
\epsffile{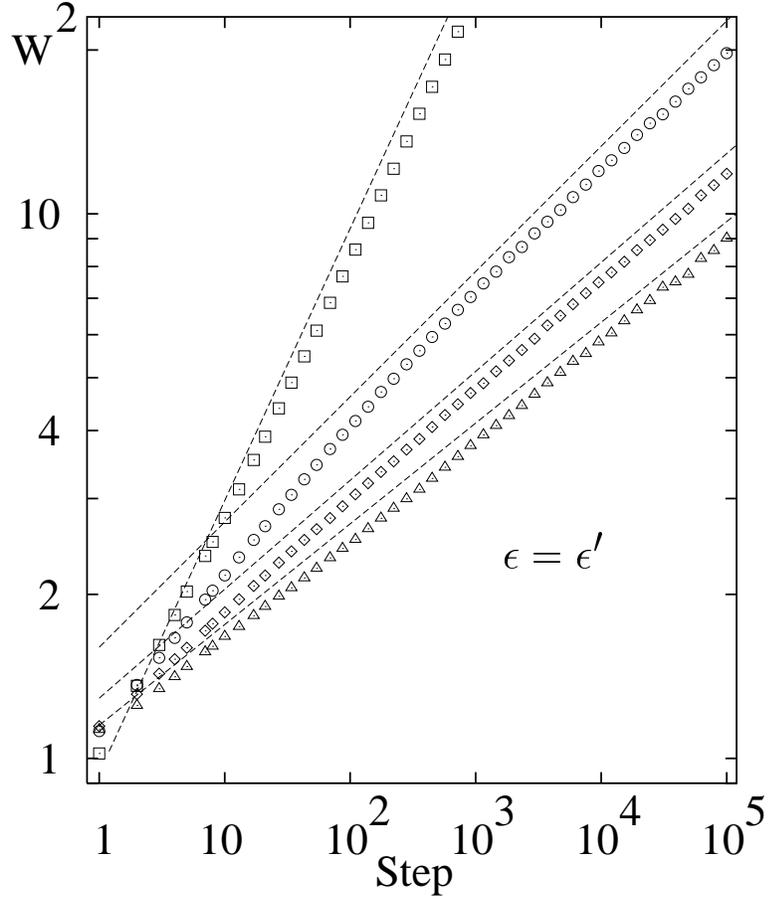}}
\vspace{-2cm}
\caption{Growth of interface width $W^2$ 
using different $k$-mers, $L= 2.1 \times 10^5$ lattice sites
and $\epsilon = \epsilon'$. For comparison, the uppermost curve 
shows the monomer case whereas 
lower curves  correspond to $k=2\,, 4,\,$ and 8,
in descending order. For $k > 1$, straight lines are fitted 
with the values of $2\beta$ given in Tab. I}

\end{figure}

\newpage
\begin{figure}
\hbox{%
\vspace{-10cm}}
\hbox{%
\hspace{-1.6cm}
\epsfxsize=8.0in
\epsffile{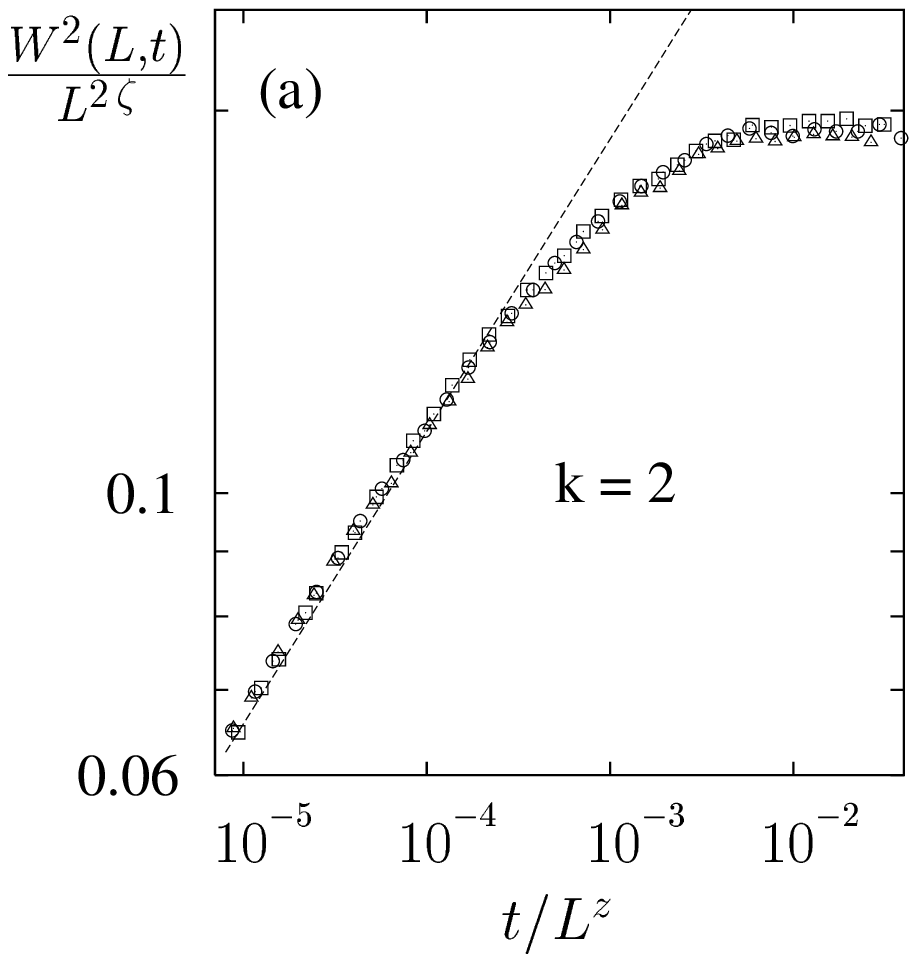}}
\vspace{-17cm}
\hbox{%
\hspace{-1.6cm}
\epsfxsize=8.0in
\epsffile{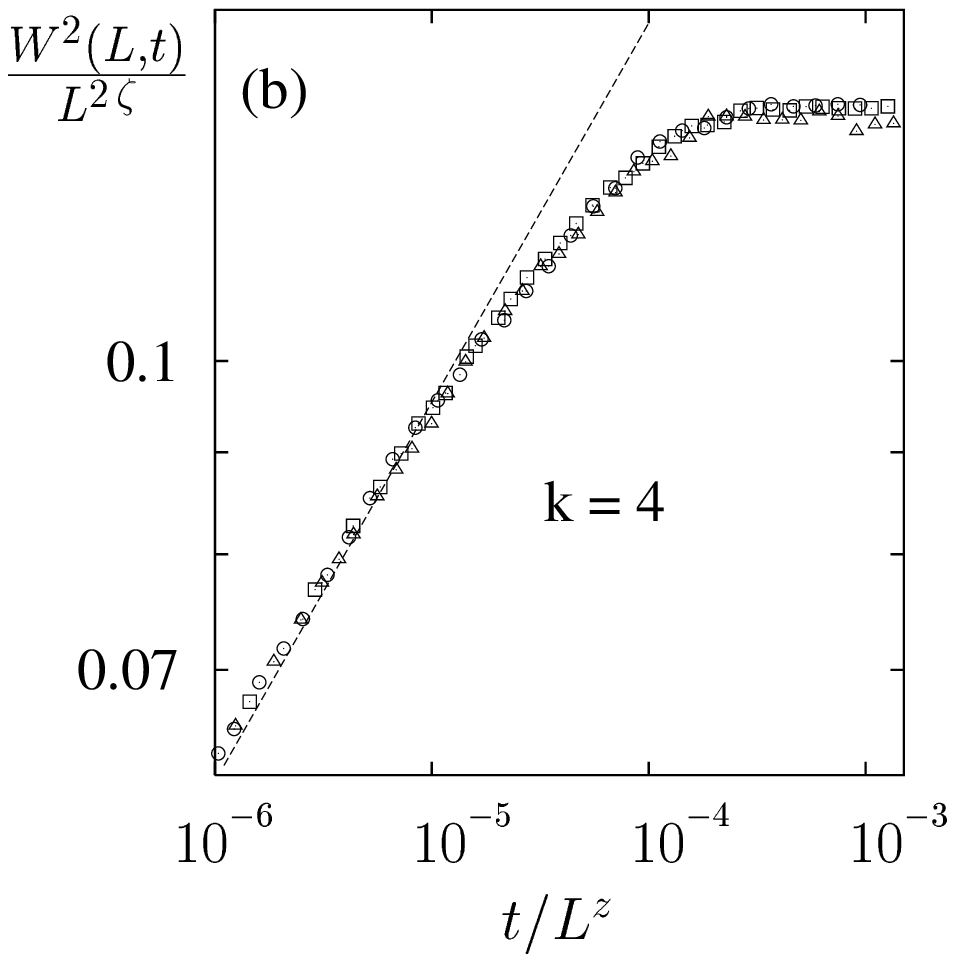}}
\end{figure}

\newpage
\begin{figure}
\vbox{%
\vspace{-10cm}}
\hbox{%
\epsfxsize=8.0in
\hspace{-1.6cm}
\epsffile{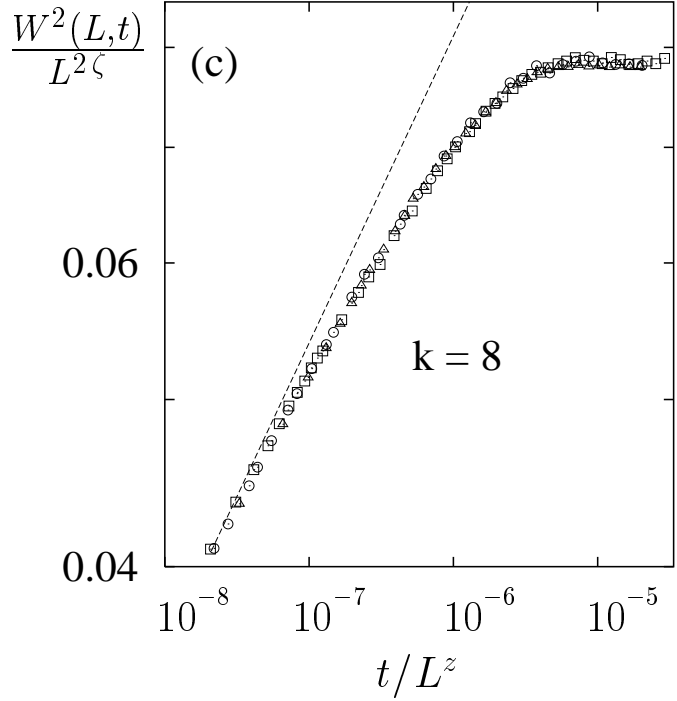}}
\vspace{-8cm}
\caption{Finite size scaling of interface width $W(L,t)$ 
using null strings and $\epsilon\! = \!\epsilon'\!$
 for (a) $k\!=\!2$, (b) $k\!=\!4$ and, (c) $k\!=\!8$. 
Sizes $L\!=\! 400,\, 640,\,$ and, 800  are
denoted respectively by squares, circles and triangles.
The data collapse was attained upon tuning the set of 
exponents $\{z,\zeta\}\,$ given in Table I. The straight 
lines are fitted with the values of $2\beta$ obtained
in the thermodynamic limit.}
\end{figure}

\newpage
\begin{figure}
\vbox{%
\vspace {-5cm}}
\hbox{%
\epsfxsize=7.5in
\hspace{-2.3cm}
\epsffile{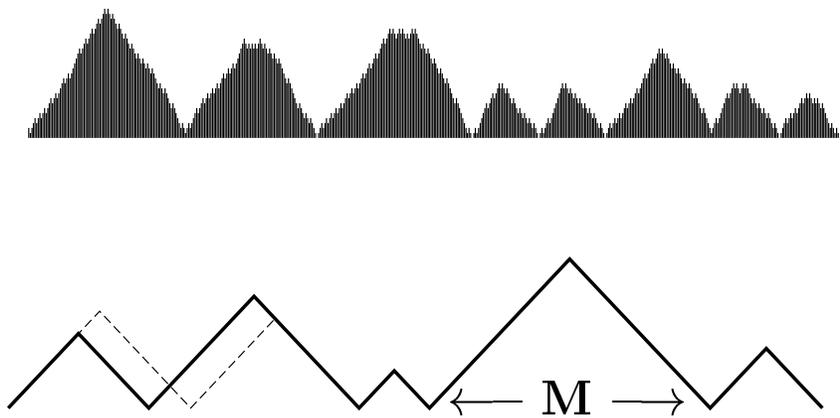}}
\vspace{-14cm}
\caption{Evolution snapshot for dimers after $10^4$ steps for 
$\epsilon'/\epsilon = 0.4$. Only 500 sites are shown (upper panel).
The dotted lines in the schematic representation denotes
the rebuilding process described in the text (lower panel). }
\end{figure}

\newpage
\begin{figure}
\vbox{%
\vspace {-4cm}}
\hbox{%
\epsfxsize=6.3in
\hspace{-1.5cm}
\epsffile{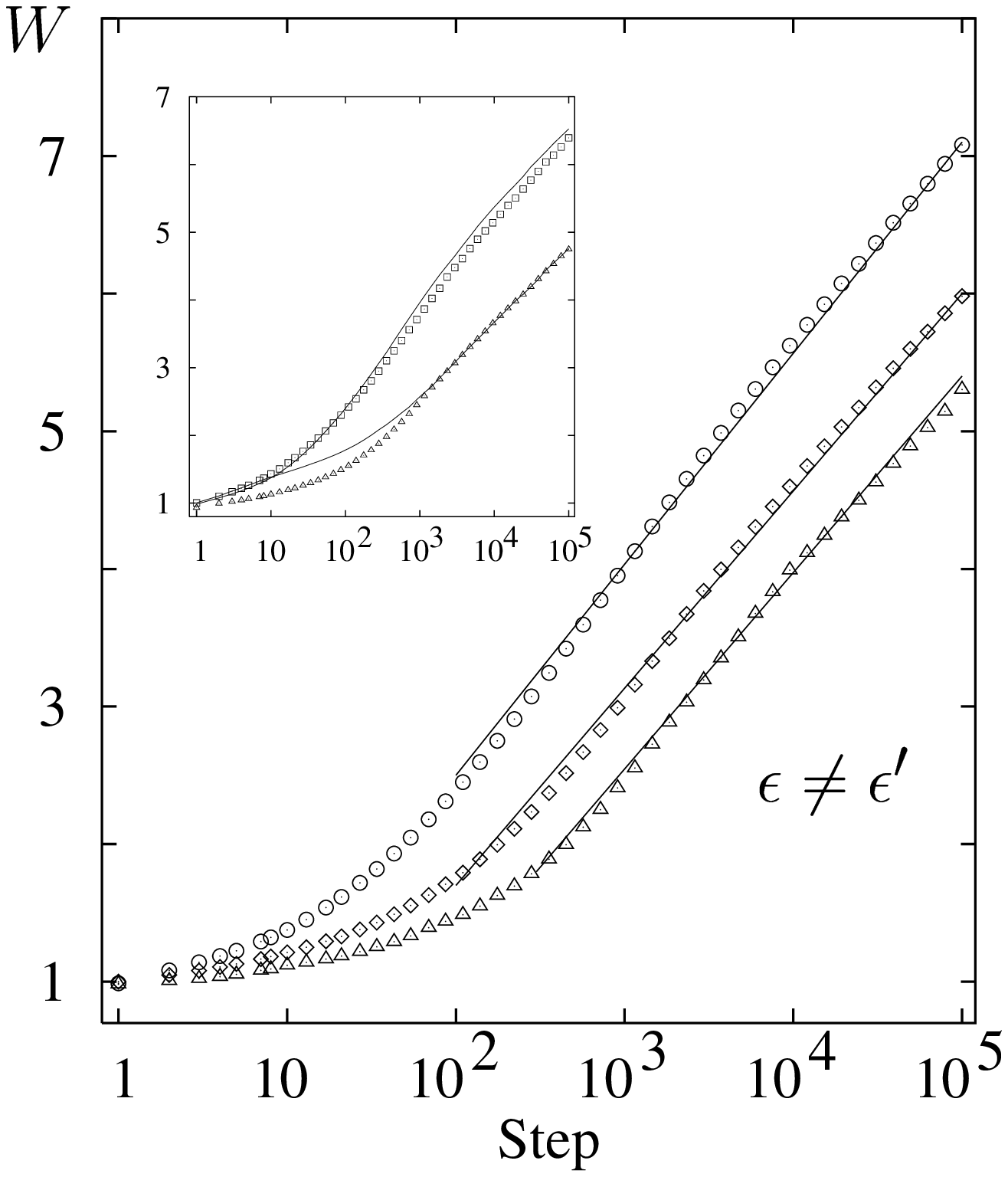}}
\vspace{-5cm}
\caption{Logarithmic growth of the interface width for 
$\epsilon'/\epsilon = 0.4\,$ starting from a flat substrate.
Straight lines over circles ($k = 2$), romboids  ($k = 4$),
and triangles ($k = 8$) are guides to the eye. The upper inset
displays similar results for $k=2$ in sectors (3) (triangles), 
and (5) (squares), which are closely and respectively followed by  
those of sectors (4) and (2) (solid lines).}
\end{figure}


\vskip 1cm
\begin{thebibliography}{99}

\bibitem{Krug} J. Krug, Adv. Phys. {\bf 46}, 139 (1997);
J. Krug and H. Spohn, in {\it Solids far from Equilibrium:
Growth, Morphology and Defects}, edited by C. Godr\`eche
(Cambridge University Press, Cambridge, 1995).

\bibitem{HZ} T. Halpin-Healy and Y.-C. Zhang, Phys. Rep.
{\bf 254}, 215 (1995).

\bibitem{Meakin} P. Meakin Phys. Rep. {\bf 235}, 189 (1993); 
P. Meakin in {\it Phase Transitions and Critical Phenomena},
edited by C. Domb and J. L. Lebowitz (Academic, New York, 1988), Vol. 12.

\bibitem{Politi} P. Politi, G. Grenet, A. Marty, A. Ponchet and J. Villain,
Phys. Rep. {\bf 324}, 271 (2000); 
D. E. Wolf and J. Villain, Europhys. Lett. {\bf 13}, 389 (1990).

\bibitem{Hwa} T. Hwa, Phys. Rev. Lett. {\bf 69}, 1552 (1992).

\bibitem{Kardar} M. Kardar and Y. -C Zhang, Phys. Rev. Lett. {\bf 58},
2087 (1987).

\bibitem{Krug2} J. Krug, Phys. Rev. Lett. {\bf 72}, 2907 (1994).

\bibitem{Newman}T. J. Newman and M. R. Swift, Phys. Rev. Lett. {\bf 79},
2261 (1997).

\bibitem{Lopez} J. M. L\'opez, Phys. Rev. Lett. {\bf 83}, 4594 (1999).

\bibitem{Koduvely} H. M. Koduvely and D. Dhar, J. Stat. Phys. {\bf 90},
57 (1998). 

\bibitem{Kertesz} J. Kert\'esz and D. E. Wolf, Phys. Rev. Lett. {\bf 62},
2571 (1989).

\bibitem{Ala-Nissila} T. Ala-Nissila, T. Hjelt, J.M. Kosterlitz
and O. Ven\"al\"ainen, J. Stat. Phys, {\bf 72}, 207 (1993).

\bibitem{Forrest} B. M. Forrest and  L.-H. Tang, Phys. Rev. Lett.
{\bf 64}, 1405 (1990).

\bibitem{Meakin2} P. Meakin, P. Ramanlal, L.M. Sander and R.C. Ball,
Phys. Rev. A {\bf 34}, 5091 (1986); M. Plischke, Z. R\'acz and 
D. Liu, Phys. Rev. B {\bf 35}, 3485 (1987).

\bibitem{Hinrichsen} Related growth process involving dimers
were introduced recently by H. Hinrichsen and G. \'Odor, Phys. Rev. Lett.
{\bf 82}, 1205 (1999); H. Hinrichsen and G. \'Odor,
Phys. Rev. E {\bf 60}, 3842 (1999);  J. D. Noh, H. Park and 
M. den Nijs, Phys. Rev. Lett. {\bf 84}, 3891 (2000).

\bibitem{Amar} J. G. Amar and F. Family, Phys. Rev. A {\bf 41}, 
3399 (1990); K. Moser and D. E. Wolf, J. Phys. A {\bf 27}, 4049 (1994).

\bibitem{KPZ} M. Kardar, G. Parisi, and Y.-C. Zhang, Phys. Rev. Lett. 
{\bf 56}, 889 (1986).

\bibitem{Barma} M. Barma and D. Dhar, Phys. Rev. Lett. {\bf 73},
2135 (1994); D. Dhar and M. Barma, Pramana {\bf 41}, L193 (1993);
M. Barma in, {\it Nonequilibrium Statistical Mechanics 
in One Dimension}, edited by V. Privman, (Cambridge University Press, 1996);
R.B. Stinchcombe, M.D. Grynberg and M. Barma, Phys. Rev. E {\bf 47},
4018 (1993).

\bibitem{Family} F. Family and T. Vicsek, J. Phys. A {\bf 18}, L75 (1985).

\bibitem{Gunter}G. M. Sch\"utz, in {\it Phase Transitions and Critical 
Phenomena}, edited by C. Domb and J. L. Lebowitz, (Academic, London 2000);  
M. D. Grynberg and R. B. Stinchcombe, Phys. Rev. E {\bf 61}, 324 (2000).


\bibitem{equilibrium} For $\epsilon = \epsilon'$,
$H$ is a self adjoint stochastic operator. Hence, the 
steady state of a given subspace is an equally weighted sum 
of all reachable configurations and therefore detailed balance holds.


\bibitem{stretched} Naturally, for large times and $\alpha > 0$
this is equivalent to the simpler form $A - B/t^{\alpha}$. However,
the best fit corresponds to the exponential form referred to above, 
implying that higher order corrections become relevant within our
accessible time scales.


\bibitem{PBC} However, PBC impose the constraint
$\sum_{n=1}^{\cal L} \chi_n = 0\,$.


\bibitem{Evans} M.R. Evans, Y. Kafri, H.M. Koduvely and D. Mukamel,
Phys. Rev. E {\bf 58}, 2764 (1998).

\end{thebibliography}
\end{document}